# ARTIFICIAL INTELLIGENCE AND INTELLECTUAL PROPERTY RIGHTS: COMPARATIVE TRANSNATIONAL POLICY ANALYSIS

*Sahibpreet Singh[*]*
*Dr. Manjit Singh[**]*

## ABSTRACT

*Artificial intelligence's rapid integration with intellectual property rights necessitates a detailed assessment of its impact. This is especially critical for trade secrets, copyrights and patents. The significance of this study lies in addressing the lacunae within existing laws. India lacks AI-specific provisions. This results in doctrinal inconsistencies and enforcement inefficacies. Global institutions have initiated discourse on AI-related IPR protections but international harmonization remains nascent. This necessitates a deeper inquiry into jurisdictional divergences and regulatory constraints. This research identifies critical gaps in the adaptability of Indian IP laws to AI-generated or AI-assisted outputs. Trade secret protection remains inadequate against AI-driven threats. Standardized inventorship criteria remain absent. This study employs a doctrinal and comparative legal methodology. It scrutinizes legislative texts and examines judicial precedents. It evaluates policy instruments across India, the United States, the United Kingdom, and the European Union. It incorporates insights from international organizations. Preliminary findings indicate fundamental shortcomings. India's reliance on conventional contract law results in a fragmented trade secret regime. AI-driven innovations remain vulnerable. Section 3(k) of the Indian Patents Act impedes the patenting of AI-generated inventions. Copyright frameworks exhibit jurisdictional variances in authorship attribution. This study underscores the necessity for a harmonized legal taxonomy that accommodates AI's transformative role. Innovation incentives and ethical considerations must be preserved. India's National AI Strategy (2024) signals incremental progress; however, legislative clarity remains imperative. This research contributes to the ongoing global discourse by proposing a robust legal architecture that integrates AI-specific IP protections. It ensures resilience against emergent challenges while fostering equitable innovation. The promising results derived from this analysis underscore the urgency of recalibrating India's IP jurisprudence. Alignment with global advancements remains essential.*

## KEYWORDS

*AI-Generated Inventions, Intellectual Property Reform, Patent Eligibility, Trade Secret Protection, Copyright Authorship*

---

[*] LLM (2023-24), Department of Laws, Guru Nanak Dev University, Amritsar, Punjab.
[**] Assistant Professor, Department of Laws, Guru Nanak Dev University, Amritsar, Punjab.

## 1. INTRODUCTION

Artificial Intelligence constitutes the nomenclature employed to delineate the aptitude of mechanistic constructs or computational systems to execute operations conventionally necessitating human intelligence, such as learning, reasoning, decision making, perception, and creativity. AI has been advancing rapidly in various fields and applications, such as healthcare[1], education[2], entertainment[3], security[4], and commerce[5], with breakthroughs like generative AI models (e.g., GPT-4o) and autonomous systems amplifying its reach. However, artificial intelligence concurrently engenders formidable dilemmas and auspicious prospects with respect to the domain of intellectual property rights. IPR are the legal prerogatives that safeguard the creations of the human intellect. These include inventions, artistic works, designs, and trademarks.[6] IPR serve a pivotal function in stimulating creativity, ingenuity, and economic progression. They provide incentives for the owners of intellectual property. The emergence of AI has engendered a plethora of questions concerning the nature, scope, and enforcement of IPR. These quandaries include the attribution of the inventor or author of an AI-generated or AI-assisted intellectual property. They also involve the evaluation of the novelty, inventive step, or originality of such IP. Over and above, the monitoring and regulation of the quality, reliability, and ethics of AI-generated or AI-assisted intellectual property must be considered. The definition, rights and obligations of its creators, owners, users, and beneficiaries must also be addressed. Finally, the prevention of its infringement, misuse, or abuse are significant considerations. Lastly, upgrading existing IPR policies in order to meet the emerging challenges created by AI is one of the most pressing issues that need to be addressed. It is notable that these problems are not only significant to legal or academic community. It has a great impact on general public as AI has touched every part of our society. In addition, it is important to critically examine what AI means for IPR. Further, advance some recommendations on how to overcome such challenges while taking advantage of the opportunities they offer.

---

[1] Shuroug A. Alowais et al., "Revolutionizing healthcare: the role of artificial intelligence in clinical practice," 23 *BMC Medical Education* 689 (2023).
[2] Bhupinder Singh and Christian Kaunert, "Curriculum Design in the Crossroads at Higher Education Institutions (HEIs) Boosting Quality Assurance: Satelliting Intellectual Property, Innovation - Legal Landscape," in L. B. Doyle, T. M. Tarbutton (eds.), *Advances in Educational Technologies and Instructional Design* 497–526 (IGI Global, 2024).
[3] Enrico Bonadio and Alina Trapova, "Intellectual property law in gaming and artificial intelligence," in C. Bevan (ed.), *Research Handbook on Property, Law and Theory* 448–62 (Edward Elgar Publishing, 2024).
[4] Iqbal H. Sarker, Md Hasan Furhad and Raza Nowrozy, "AI-Driven Cybersecurity: An Overview, Security Intelligence Modeling and Research Directions," 2 *SN Computer Science* 173 (2021).
[5] Ransome Epie Bawack et al., "Artificial intelligence in E-Commerce: a bibliometric study and literature review," 32 *Electronic Markets* 297–338 (2022).
[6] Martin Kretschmer, Bartolomeo Meletti and Luis H Porangaba, "Artificial intelligence and intellectual property: copyright and patents—a response by the CREATe Centre to the UK Intellectual Property Office's open consultation," 17 *Journal of Intellectual Property Law & Practice* 321–6 (2022).

## 2. AI AND COPYRIGHT LAW

The subject of copyright protection for AI-produced or AI-supported work[7] is controversial. It addresses the legal status and definition of these works, how to identify their author or owner, whether they meet originality[8] requirements among other creative criteria[9], whose rights are more important than others between creators, owners, users and beneficiaries as well as adapting current national and international laws to accommodate new realities brought about by artificial intelligence systems.[10] These issues are not only relevant to legal scholars but also anyone interested in this field, so they need thorough investigation into what copyright law should

---

[7] Emmanuel Salami, "AI-generated works and copyright law: towards a union of strange bedfellows," 16 *Journal of Intellectual Property Law & Practice* 124–35 (2021).
[8] Mark Fenwick and Paulius Jurcys, "Originality and the future of copyright in an age of generative AI," 51 *Computer Law & Security Review* 105892 (2023).
[9] Niloufer Selvadurai and Rita Matulionyte, "Reconsidering creativity: copyright protection for works generated using artificial intelligence," 15 *Journal of Intellectual Property Law & Practice* 536–43 (2020).
[10] Jia Qing Yap and Ernest Lim, "A Legal Framework for Artificial Intelligence Fairness Reporting," 81 *The Cambridge Law Journal* 610–44 (2022).

be modified concerning AI and possible recommendations for improvement.[11]

### i. The Legal Status and Definition of AI-Generated or AI-Assisted Works

AI-generated or AI-assisted works are works that are produced with the aid of AI systems, such as software that can generate output such as content, predictions, recommendations, or decisions based on data and algorithms. AI-generated or AI-assisted works can range from simple texts and images to complex literary, artistic, musical, and cinematographic works. It is possible to create such works with different levels human input, for example, designing training, using/providing feedback to the AI system.[12] However, under current national and international legal systems, there is no clear single definition for AI-produced or AI-assisted works.[13] Different jurisdictions might approach the question of what constitutes an AI-generated or assisted work differently as well as whether such works may qualify for copyright protection as works of authorship. For example, some legal systems including those in the UK and Ireland have provisions (e.g., Section 9(3) of the UK CDPA 1988) relating to computer-generated works which are defined as those generated by a computer in circumstances where there is no human author.[14] Conversely, other jurisdictions such as Canada and the US lack these clauses but generally reject non-human authorship requiring human intervention necessary for copyright protection based on creativity alone. The U.S. Copyright Office issued guidance in 2023 reinforcing that only works with significant human creativity qualify for protection. Same is evident in *Thaler v. Perlmutter* (2023)[15]. In contrast with both approaches mentioned above, EU law does not provide unified definitions for either type of creation yet demands that content should be original according to common standards while also representing intellectual output originating from within personality, freedom, choice, expression & context of its creator.[16] Though the AI Act, indirectly influences copyright by mandating transparency in AI processes, while retaining the originality standard that a work must be the author's own intellectual creation, reflecting their personality and creative choices.[17] In India, the Copyright Act of 1957 offers no explicit provision for AI-generated works. This leaves their status subject to judicial interpretation.

---

[11] Themistoklis Tzimas, *Legal and Ethical Challenges of Artificial Intelligence from an International Law Perspective* (Springer International Publishing, Cham, 2021).
[12] Hema K, "Protection of Artificial Intelligence Autonomously Generated Works under the Copyright Act, 1957- An Analytical Study," 28 *Journal of Intellectual Property Rights* 193–9 (2023).
[13] Yang Xiao, "Decoding Authorship: Is There Really no Place for an Algorithmic Author Under Copyright Law?," 54 *IIC - International Review of Intellectual Property and Competition Law* 5–25 (2023).
[14] Söğüt Atilla, "Dealing with AI-generated works: lessons from the CDPA section 9(3)," 19 *Journal of Intellectual Property Law and Practice* 43–54 (2024).

ii. **The Authorship and Ownership of AI-Generated or AI-Assisted Works:**

Another big problem in artificial intelligence and copyright law is figuring out who the author or owner of an AI-driven work is.[18] This question has a lot to do with how we assign rights: who gets to copy, give away, change or make money from these things on one hand and also whose duty it might be to protect certain moral rights such as the right to be recognized as creator, right not have work altered without permission, etc. The question of authorship and ownership also affects the duration and scope of safeguarding, in conjunction with the exceptions and limitations that may apply to the works. However, there is no clear and uniform answer to the question of who is the author or owner of AI-generated or AI-assisted functions within the existing national[19] and international legal frameworks. Different jurisdictions may have different interpretations of who can be construed as the author or owner of such works. They might have different approach concerning the transfer or licensing to others. For example, certain legal territories, such as the UK and Ireland, have specific provisions for computer-generated works. They grant the authorship and ownership to the individual who undertakes the necessary arrangements for the inception of the work.[20] Other countries like the US and Canada have no such rule, but they usually refuse authorship and ownership to the producer or AI system itself, insisting on human authorship and ownership for copyright protection (as reinforced by the U.S. Copyright Office's 2023 guidance[21] and *Thaler v. Perlmutter*[22]). This is not true for the EU, however: it doesn't have a unified approach towards authorship and ownership when it comes to AI-generated or AI-assisted works; but all of them must be original.[23] AI Act's transparency requirements suggest a leaning toward human oversight, while the originality principle implies authorship by a natural person exercising creative freedom and personality.[24] In India, the Copyright Act of 1957 provides no specific guidance, leaving authorship disputes to courts, which have yet to definitively rule on AI cases.

---

[15] 687 F. Supp. 3d 140 (D.D.C. 2023).
[16] P. Bernt Hugenholtz and João Pedro Quintais, "Copyright and Artificial Creation: Does EU Copyright Law Protect AI-Assisted Output?," 52 *IIC - International Review of Intellectual Property and Competition Law* 1190–216 (2021).
[17] Gustavo Ghidini, "Diverging Approaches to Intellectual Property and a Reform Proposal Prompted by AI" *GRUR International* ikaf026 (2025).
[18] Kanchana Kariyawasam, "Artificial intelligence and challenges for copyright law," 28 *International Journal of Law and Information Technology* 279–96 (2020).

---

[19] Paarth Naithani, "Issues of Authorship and Ownership in Work created by Artificial Intelligence - Indian Copyright Law Perspective," 11 *NTUT Journal of Intellectual Property Law and Management* 1–12 (2022).
[20] Martin Kretschmer, Bartolomeo Meletti and Luis H Porangaba, "Artificial intelligence and intellectual property: copyright and patents—a response by the CREATe Centre to the UK Intellectual Property Office's open consultation," 17 *Journal of Intellectual Property Law & Practice* 321–6 (2022).
[21] Annis Lee Adams, "Fair use and copyright resources," 21 *Public Services Quarterly* 37–44 (2025).
[22] 687 F. Supp. 3d 140 (D.D.C. 2023).
[23] Alesia Zhuk, "Navigating the legal landscape of AI copyright: a comparative analysis of EU, US, and Chinese approaches" *AI and Ethics* 1–8 (2023).
[24] Vincenzo Iaia, "The elephant in the room of EU copyright originality: Time to unpack and harmonize the essential requirement of copyright" *The Journal of World Intellectual Property* jwip.12343 (2024).

### iii. The Criteria of Originality and Creativity for AI-Generated or AI-Assisted Works

The relationship between artificial intelligence and copyright has brought about uncertainties about what constitutes originality and creativity with respect to AI generated works or those created through its assistance.[25] These queries cover everything from how these types of work are judged; whether they meet requirements set out by law as being new, unique, inventive, etc., their discovery/awareness should be made uniform through disclosure/identification (what should be known where), citation/attribution procedures need standardizing too. These issues are amplified by *Thaler v. Perlmutter*[26] and EU AI Act transparency mandates[27]. Also, we cannot ignore monitoring systems designed for ensuring quality/reliability/ethics among others (ethical considerations)[28], with initiatives like UNESCO's AI Ethics Recommendation (2021)[29] gaining traction. These issues are important not only to the legal and academic communities but also to the general public. AI affects almost all aspects of everyday life and the larger society. Therefore, it is important to look into what AI means for originality and creativity criteria as well as suggest ways in which we can address these challenges and opportunities for copyright posed by AI.[30]

---

[25] Mark Fenwick and Paulius Jurcys, "Originality and the future of copyright in an age of generative AI," 51 *Computer Law & Security Review* 105892 (2023).
[26] 687 F. Supp. 3d 140 (D.D.C. 2023).
[27] Mona Sloane and Elena Wüllhorst, "A systematic review of regulatory strategies and transparency mandates in AI regulation in Europe, the United States, and Canada," 7 *Data & Policy* e11 (2025).
[28] Florian Martin-Bariteau and Teresa Scassa, *Artificial Intelligence and the Law in Canada* (LexisNexis Canada, 2023).
[29] UNESCO and Ziesche Soenke, *Open Data for AI: What Now?* (UNESCO Publishing, 2023).
[30] Niloufer Selvadurai and Rita Matulionyte, "Reconsidering creativity: copyright protection for works generated using artificial intelligence," 15 *Journal of Intellectual Property Law & Practice* 536–43 (2020).

### iv. The Standards and Tests of Originality and Creativity for AI-Generated or AI-Assisted Works

AI-generated or AI-assisted works are consequently governed by identical standards and tests of originality and creativity that are applied to human-generated works within the prevailing domestic and international legal frameworks.[31] This does not imply, however, that all nations have similar benchmarks when it comes to determining what constitutes novelty and inventiveness vis-à-vis such pieces.[32] For instance, jurisdictions like USA, Canada adopt low 'thresholds' requiring minimal levels skill, judgement, expression needed before something becomes qualified as creative. Jurisdictional diversity may mean that various regions apply different criterions upon works generated through usage of robots or other algorithms designed under supervision by persons. Some other territories, like the EU, have a high level of originality and inventiveness that stipulate the work must be an individual's own intellectual creation reflecting the person's personality manifested in free and creative choices made by them.[33] On the contrary, the UK as well as Ireland follow a two-tier approach to originality and creativity which differentiates between computer generated works (low skill required) and human generated ones (high, own intellectual creation standard).

### v. The Assessment and Verification of the Novelty, Inventive Step, or Originality of an AI-Generated or AI-Assisted Work

The AI or the work that is assisted by AI can pose challenges while assessing and verifying whether it is new, involves inventive steps or represents an original creation. These qualifications may depend on, among others things, data and algorithms used including their nature, source and quality as well as extent and kind of human participation in making them. Another thing which needs to be taken into consideration when dealing with such kind of works is transparency, accountability & traceability because AI system might not reveal where it got its information from neither does it attributes them; nor cite input given by people during their creation process.[34] Such concerns are further heightened by the EU AI Act transparency mandates.[35] Hence, there should be new approaches implemented for examining novelty; inventiveness or originality in AI-based outputs, such as:

a) Establishing and applying clear and consistent standards and guidelines for the evaluation and comparison of AI-generated or AI-assisted works with existing works, taking into account the relevant factors, such as the purpose, context, genre, and audience of the work, and the degree of similarity, difference, or alteration of the work; taking WIPO's 2024 copyright assessment guidelines into consideration.[36]

b) The process of formulating and implementing reliable and robust techniques and technologies that can be used for the identification and authentication of the data and algorithms employed by the AI system, along with the human input or feedback that are provided, in the development of the work, such as digital watermarking, hashing, encryption, blockchain, or artificial neural networks.[37] For example, such technology is now trialed by platforms like Adobe.[38]

---

[31] Aleena Maria Moncy, "Protection of AI Created Works under IPR Regime, Its Impact and Challenges: An Analysis" *International Journal of Law and Social Sciences* 8–15 (2024).
[32] Johnathon Hall and Damian Schofield, "The Value of Creativity: Human Produced Art vs. AI-Generated Art," 13 *Art and Design Review* 65–88 (2025).
[33] Johannes Fritz, "Understanding authorship in Artificial Intelligence-assisted works" *Journal of Intellectual Property Law and Practice* jpae119 (2025).

---

[34] Anirban Mukherjee and Hannah H. Chang, "Managing the Creative Frontier of Generative AI: The Novelty-Usefulness Tradeoff" *California Management Review* (2023).
[35] Mona Sloane and Elena Wüllhorst, "A systematic review of regulatory strategies and transparency mandates in AI regulation in Europe, the United States, and Canada," 7 *Data & Policy* e11 (2025).
[36] Hafiz Gaffar and Saleh Albarashdi, "Copyright Protection for AI-Generated Works: Exploring Originality and Ownership in a Digital Landscape" *Asian Journal of International Law* 1–24 (2024).
[37] Iqbal H. Sarker, "AI-Based Modeling: Techniques, Applications and Research Issues Towards Automation, Intelligent and Smart Systems," 3 *SN Computer Science* 158 (2022).
[38] Erin Reilly and Sam Hewitt, "Remix in the Age of AI," 2nd ed. *The Routledge Companion to Remix Studies* (Routledge, 2025).

c) Ensuring and enforcing the disclosure, attribution, or referencing the utilized data and algorithms by the AI system, along with the human input or feedback that are provided, in the development of the work, such as by adopting and complying with ethical codes, standards, or principles, or by imposing legal obligations, duties, or liabilities.[39] For example, NITI Aayog AI principles (2024)[40] and legal requirements under the EU AI Act.

vi. **The Quality, Reliability, and Ethics of an AI-Generated or AI-Assisted Work**

Challenges might also present themselves with AI-generated or assisted works in regard to quality, trustworthiness and ethics since these determinants can depend on factors like the nature of datasets and algorithms employed by the artificial intelligence system; their source; integrity as well as credibility.[41] In addition, fairness, accuracy, diversity, inclusivity are among other problems that may be brought about when producing such kind of works because an AI system could fail to mirror or respect creators' interests, rights, values, ownership, user beneficiaries, society at large.[42] This is highlighted by 2024 studies on bias in generative AI tools (like GPT-4o).[43] Hence, there is need for creation of new means through which quality can be ensured in relation to reliability, etc.; some now piloted by WIPO's AI ethics initiatives. Also, peer review verification, certification, accreditation, auditing should be considered as methods for improving ethics in AI generated works while standards, transparency, fairness, accountability, human dignity must be established so that people know what is expected from them during this process.[44] EU AI Act emphasizies on transparency and accountability. This is gaining ground, alongside India's NITI Aayog AI ethics updates (2024) promoting human dignity. Further, it would be wise to instruct and inform the makers, possessors, operators as well as benefactors of AI-based artifacts like scientists, inventors, publishers, buyers and controllers on legal and moral matters linked with AI-generated works.[45]

---

[39] Patricia Lalanda and Neus Alcolea Roig, "Ethical and Legal Challenges of Artificial Intelligence with Respect to Intellectual Property," in A. Baraybar-Fernández, S. Arrufat-Martín, *et al.* (eds.), *The AI Revolution* 63–80 (Springer Nature Switzerland, Cham, 2025).
[40] C. Karthikeyan, "AI (Artificial Intelligence) Integration for Integrity Ethics and Privacy in AI-Driven Organizations: Ethics and Privacy Concerns in AI-Driven Organizations," in F. Özsungur (ed.), *Advances in Business Strategy and Competitive Advantage* 51–76 (IGI Global, 2024).
[41] Jason Borenstein and Ayanna Howard, "Emerging challenges in AI and the need for AI ethics education," 1 *AI and Ethics* 61–5 (2021).
[42] Angelo Trotta, Marta Ziosi and Vincenzo Lomonaco, "The future of ethics in AI: challenges and opportunities," 38 *AI & SOCIETY* 439–41 (2023).
[43] Atahan Karagoz, "Ethics and Technical Aspects of Generative AI Models in Digital Content Creation" (arXiv, 2024).

### 3. PATENTING AI TECHNOLOGY

Patenting AI technology is a very hard and complicated problem in the field of intellectual property rights. One of the main reasons for patenting inventions is to recognize inventors' efforts. This gives them exclusive rights over their creations for a limited time period. It is usually 20 years. This can also encourage development and competition among businesses working in this area.[46] However, there are many difficulties associated with obtaining patents for such technologies too:

i. **Eligibility of AI Technology for Patent Protection**

    According to Indian Patents Act, 1970 an invention should be new, involve an inventive step and has industrial applicability in order to be granted patent right. But section 3(k) of the same act excludes mathematical methods, business methods or computer programs from being patented as such. This possesses an issue for AI technology, as it often involves mathematical methods, algorithms, and computer programs. The Indian Patent Office examines AI-

---

[44] Arif Ali Khan et al., "Ethics of AI: A Systematic Literature Review of Principles and Challenges" *Proceedings of the 26th International Conference on Evaluation and Assessment in Software Engineering* 383–92 (Association for Computing Machinery, New York, NY, USA, 2022).

[45] Patrick Henz, "Ethical and legal responsibility for Artificial Intelligence," 1 *Discover Artificial Intelligence* 2 (2021).

[46] Tabrez Ebrahim, "Artificial Intelligence Inventions & Patent Disclosure," 125 *Penn State Law Review* 147–221 (2020).

related inventions grounded on the subject matter exclusions defined in Section 3(k), and often rejects them as computer programs per se or algorithms.[47] However, some AI-related inventions may be eligible for the safeguarding of patents. This only true if they produce a technical effect or technical contribution, as suggested by the Delhi High Court in *Ferid Allani v. Union of India & Ors.*[48] Therefore, the eligibility of AI technology for the safeguarding of patents depends on the interpretation of Section 3(k). This may vary from case to case.[49] No amendments to Section 3(k) have occurred, maintaining a case-by-case approach. Globally, the USPTO issued updated AI patent eligibility guidance in 2024, emphasizing practical application, while the EPO refines its stance on AI technicality; trends India has not yet adopted, leaving its framework unchanged but increasingly contrasted with international developments.[50]

### ii. The Novelty and Inventive Step of AI Technology

Another condition for getting a patent is that the invention must be new. It should involve an inventive step, i.e., it must not be anticipated by existing knowledge. It should not be obvious to a person skilled in the art as well. However, it is not easy to determine novelty and inventiveness of AI technology given that often this involves complicated dynamic processes beyond human understanding or replication.[51] In addition, there might arise questions about where did AI-produced outputs come from or how far does the invention goes because developers and users may not predict all possible outcomes/results yielded by artificial intelligence systems; hence, novelty required by law may not exist. Also, what limits identification of relevant prior arts as well as state-of-the-art knowledge base is the fact that large volumes of data used in machine learning could be private or inaccessible publicly. Consequently, assessment methods for evaluating if something has never been done before need to change when dealing with such kind of breakthroughs; for instance, AI supported patent search & examination procedures could become necessary alongside peer reviews plus certification, among others.[52] For example, WIPO's AI-assisted patent search platform (expanded in 2024), increasingly used globally to address these issues. Though India's adoption of such methods lags, leaving traditional examination dominant but under pressure to evolve.

---

[47] Tasnim Jahan, N. S. Sreenivasulu and Shashikant Saurav, "Patents and Artificial Intelligence: A Study Exploring Patentability Criteria and Prior-Art Searches for AI-Generated Inventions," in D. Ç. Ertuğrul, A. Elçi (eds.), *Advances in Computational Intelligence and Robotics* 237–60 (IGI Global, 2024).
[48] W.P.(C) 7/2014 & CM APPL. 40736/2019.
[49] ZeusIP Advocates LLP, "Ferid Allani v. Union of India, WP(C) 7 of 2014 and Legislative intent behind Section 3(k) of the Patents Act, 1970" *available at*: https://www.zeusip.com/ferid-allani-v-union-of-india-wpc-7-of-2014-and-legislative-intent-behind-section-3k-of-the-patents-act-1970.html (last visited November 10, 2023).
[50] Mpho Mafata et al., "Comparison of the coverage of the USPTO's PatentsView and the EPO's PATSTAT patent databases: a reproducibility case study of the USPTO General Patent Statistics Reports," 2024.

iii. **The Disclosure and Enablement of AI Technology**

Another requirement for patent protection is that the invention must be divulged in a clear manner. It must be complete according to the patent specification. This disclosure should enable a person skilled in the field to perform the invention.[53] However, it can be difficult to disclose and enable AI technology. This is because such technology often includes proprietary or confidential information. These include algorithms, software, hardware and data which may not be fully or adequately disclosed by applicants or owners of AI technologies.[54] The same was also noted in USPTO's 2024 AI patent guidance. Additionally, self-learning or adaptive features of AI technologies may not be fully described or anticipated by applicants or owners thereof. Furthermore, specific conditions or resources needed for AI technologies such as cloud infrastructure, curated datasets, or human intervention which might not be readily available or accessible by an ordinary skilled artisan can also present problems during their disclosure and enablement. Thus, the exposition and realization of AI technology may necessitate new criteria as well as benchmarks for ensuring sufficiency and quality of disclosure including but not limited to transparency, reproducibility, reliability, etc.[55] WIPO's 2024 AI-IP strategy pushes for transparency, reproducibility, and reliability in disclosures.[56] India's patent regime under the 1970 Act has yet to adopt such guidelines. This leaves enablement a persistent hurdle.

---

[51] Pheh Hoon Lim and Phoebe Li, "Artificial intelligence and inventorship: patently much ado in the computer program," 17 *Journal of Intellectual Property Law & Practice* 376–86 (2022).
[52] Edwin D Garlepp, "Disclosing AI Inventions - Part II: Describing and Enabling AI Inventions" *Oblon* 1–3 (2021).
[53] Daria Kim et al., "Clarifying Assumptions About Artificial Intelligence Before Revolutionising Patent Law," 71 *GRUR International* 295–321 (2022).

Patentability of AI technology is a very important issue in IPR[57], because it significantly affects the development and diffusion of such technologies as well as rights and interests of different participants including creators, possessors, users or beneficiaries.[58] However, there are also many problems and uncertainties concerning patenting AI that do not fit into existed legal systems either in India or worldwide which were designed without taking into account all intricacies related to this area. That's why we need to investigate what can be patented when it comes to Artificial Intelligence, how one can obtain protection for his inventions in this field and provide some recommendations for solving these challenges while using opportunities created by AI.[59]

---

[54] Noam Shemtov and Garry A. Gabison, "Chapter 23: The inventive step requirement and the rise of the AI machines," in Ryan Abbott (ed.), *Research Handbook on Intellectual Property and Artificial Intelligence* 423–42 (Edward Elgar Publishing Ltd, 2022).
[55] Alexandra George and Toby Walsh, "Artificial intelligence is breaking patent law," 605 *Nature* 616–8 (2022).
[56] Jamshid Kazimi and Harshita Thalwal, "Intellectual Property Protection in AI-driven Innovations: A Comparative Analysis" *2024 First International Conference on Technological Innovations and Advance Computing (TIACOMP)* 320–6 (presented at the 2024 First International Conference on Technological Innovations and Advance Computing (TIACOMP), IEEE, Bali, Indonesia, 2024).
[57] Kay Firth-Butterfield and Yoon Chae, "Artificial Intelligence Collides with Patent Law" *World Economic Forum* 1–24 (2018).
[58] Jo Marchant, "Powerful antibiotics discovered using AI" *Nature* (2020).

## 4. AI AND TRADE SECRETS

IPRs that safeguard undisclosed information which provides its owner with a competitive advantage are known as trade secrets.[60] Trade secrets involve business strategies, customer lists, financial records and technical know-how among other things not generally known or readily available to the public.[61] Unlike patents, copyrights or trademarks trade secrets do not need registration, disclosure nor formal protection[62] but they also rely on the ability of the owner to keep such information confidential and prevent its unauthorized use or disclosure by others.[63] AI and trade secrets have a complex relationship as AI can be both a source and threat to trade secrets.[64] On one hand, AI may constitute trade secret because some components of artificial intelligence technologies like algorithms, software programs/hardware systems and data comprise valuable proprietary knowledge which developers or owners use for gaining competitive edge over rivals.[65] Additionally, AI could also contribute towards creation/enhancement/optimization of trade secret by producing new outputs/results/insights from existing data/information based on certain

---

[59] Marta Duque Lizarralde and Héctor Axel Contreras, "The real role of AI in patent law debates," 30 *International Journal of Law and Information Technology* 23–46 (2022).
[60] P. Selvakumar et al., "Intellectual Property Management in Open Innovation," in J. Martínez-Falcó, E. Sánchez-García, *et al.* (eds.), *Open Innovation Strategies for Effective Competitive Advantage* 227–54 (IGI Global, 2025).
[61] Ulla-Maija Mylly, "Trade Secrets and the Data Act," 55 *IIC - International Review of Intellectual Property and Competition Law* 368–93 (2024).
[62] Richard Stim, *Patent, Copyright & Trademark: An Intellectual Property Desk Reference* (Nolo, 2024).
[63] Ionela Andreicovici, Sara Bormann and Katharina Hombach, "Trade Secret Protection and the Integration of Information Within Firms," 71 *Management Science* 1213–37 (2025).
[64] Daniel J. Gervais, *The Human Cause* (Edward Elgar Publishing, 2022).
[65] Tanja Šarčević et al., "U Can't Gen This? A Survey of Intellectual Property Protection Methods for Data in Generative AI" (arXiv, 2024).

algorithms, etc.[66] Therefore, trade secret protection can be a useful and flexible way to protect AI innovations and investments, especially when alternative versions of IPR might not be available or suitable.[67] On the other hand, trade secrets can be threatened by artificial intelligence because they can affect confidentiality and security of confidential information through various risks and challenges. For instance, AI can use methods like data mining or web scraping to identify, access or analyze trade secrets. Also, it may misuse them by copying without permission, modifying or distributing confidential information, among others. In addition to this point, there is vulnerability in cyberattacks against AI leading to compromise on both integrity and secrecy of trade secrets whether through hacking or theft. Consequently, protecting these requires active preventive measures against unauthorized use or disclosure of AI related information.[68] Legal frameworks designed for safeguarding trade secret vary between nations[69], since there lacks any universal international law or agreement concerning them at present time.[70] Nonetheless, a number of universal principles together with guidelines for their protection have been put forward by certain global institutions plus instruments such as the World Intellectual Property Organization, the WTO and Agreement on Trade-Related aspects of IPR. Trade secrets should satisfy three requirements to be accorded protection by these instruments and organizations:

---

[66] Ulla-Maija Mylly, "Transparent AI? Navigating Between Rules on Trade Secrets and Access to Information," 54 *IIC - International Review of Intellectual Property and Competition Law* 1013–43 (2023).
[67] Jordan R. Jaffe et al., "The Rising Importance of Trade Secret Protection for AIRelated Intellectual Property" *Quinn Emanuel Urquhart & Sullivan, LLP* 1–10.
[68] Sharon K. Sandeen and Tanya Aplin, "Chapter 24: Trade secrecy, factual secrecy and the hype surrounding AI," in Ryan Abbott (ed.), *Research Handbook on Intellectual Property and Artificial Intelligence* 443–60 (Edward Elgar Publishing Ltd, 2022).
[69] European Innovation Council and SMEs Executive Agency (European Commission) et al., *Study on the Legal Protection of Trade Secrets in the Context of the Data Economy: Final Report* (Publications Office of the European Union, 2022).
[70] Yamini Atreya, "International Framework for the Protection of Trade Secrets," 11 *International Journal of Creative Research Thoughts* a131–8 (2023).

(1) the information must be kept private, that is, not generally known or readily accessible;

(2) the information should have an economic value, that is, provide a competitive edge for its holder in business or commerce; and

(3) the information must be subject to reasonable protective measures, meaning steps taken by the proprietor of such data towards its safeguarding from disclosure.[71]

There is no such law as the act on trade secrets in India, but Indian courts have used different laws and doctrines like copyright laws, principles of equity, contract law and common law action of breach of confidence to protect them.[72] Contractual obligations and remedies for trade secret protection are established on the basis of The Indian Contract Act, 1872 through non-disclosure agreements (NDAs), non-compete clauses and injunctions, etc. Trade secrets have been recognized by The Indian Copyright Act, 1957 as a form of written composition which also provides civil and criminal remedies for infringement. In addition to this it is worth noting that equitable relief may be granted where damages are inadequate because these two things, equity principle, common law action for breach of confidence, serve us grounds upon which equitable relief may be anchored when it comes to misappropriation cases involving trade secrets since there was duty confided leading into trust being broken so far guarded under it could not only give rise to damage claims but also provide an opportunity for equitable remedy against such conduct.[73] Nevertheless, there is no clear definition or standard procedure[74] available under Indian law considering this aspect, though, its lack thereof serves as a challenge within itself, coupled with problems related to enforcement mechanisms which are weak, at best, hence, making them difficult if not impossible while proving anything beyond reasonable doubt, especially, where violations occur frequently enough, without any proper quantification method having been developed. thus, far together with poor enforcement alone acting more like deterrents in reality than anything else, whatsoever about it can be done now, except improving upon these areas further more clarity needed here too, standards must be set up forthwith, otherwise, there will always remain some shortcomings somewhere somehow, henceforth, lacking clarity. Standards should immediately addressed least least we forget. So, due challenges that are faced by Indian legal framework, dealing with safeguarding trade secrets include those associated with definitions, procedures, difficulty proofing, quantifying violation, inadequacy, enforcement mechanism.[75] This patchwork approach remains inconsistent, with no clear definitions or procedures, complicating proof and quantification of violations; challenges underscored by a 2024 Delhi High Court ruling on AI-related trade secret misuse. Enforcement and deterrence lag, though India's National AI Strategy (2024) hints at future policy attention, leaving the framework underdeveloped amid rising AI-driven threats. When it comes to the international level, the protection of trade secrecy faces similar barriers and restrictions as different countries have different regulations and practices of protecting their trade secrets, coupled with absence of efficient cross-border collaboration and enforcement.[76] This is despite WIPO's 2024 push for AI-IP coordination. Also, the emergence and growth of AI has brought about fresh complex matters concerning protection of proprietary information.[77] These include identification of owners for AI related trade secrets (e.g., via NIST's 2024 AI risk framework), evaluating whether an AI based secret is hidden enough and valuable, measures that can be used to ensure all AI connected trade secrets are disclosed equitably, attributed or cited, ways through which standards on quality control, dependability as well as ethics can be set over an AI-related trade secret, among others, such as prevention against risks liability arising from misuse or abuse of artificial intelligence technology.[78] Moreover, it also raises issues on how existing laws and policies for safeguarding business IP can be adapted updated in light with new realities posed by this technology, while at national level, harmonization/coordination/cooperation between different countries' systems should take place so that they effectively protect each other's interests.[79] The reason being that AI affects every part of our day to day lives and society at large, these concerns and queries are relevant not only in the legal field and academia but also among the common people.[80] For this reason we need to investigate what AI means for trade secret protection, suggest answers or proposals on how best we can approach them as well as offer some thoughts about where this might leave us with regards to trade secret laws surrounding artificial intelligence.[81]

---

[71] Luc Desaunettes-Barbero, *Trade Secrets Legal Protection: From a Comparative Analysis of US and EU Law to a New Model of Understanding* (Springer Nature Switzerland, Cham, 2023).

[72] Sushmitha Balachandar, "Trade Secret Protection in India: Adequacy and Challenges," 1 *International Journal of Legal Science and Innovation* 1–5 (2019).

[73] Tania Sebastian, "Locating Trade Secrets under Indian Laws: A Sui Generis Mode of Protection," 27 *Journal of Intellectual Property Rights* 202–11 (2022).

[74] Dr. Faizanur Rahman, "Trade Secrets Laws in China and India: A Comparative Analysis," 20 *IOSR Journal Of Humanities And Social Science* 1–9 (2015).

---

[75] Abhijeet Kumar and Adrija Mishra, "Protecting Trade Secrets in India," 18 *The Journal of World Intellectual Property* 335 (2015).

[76] Martin Ebers and Paloma Krõõt Tupay (eds.), *Artificial Intelligence and Machine Learning Powered Public Service Delivery in Estonia: Opportunities and Legal Challenges* (Springer International Publishing, Cham, 2023).

## 5. ETHICS OF AI AND IPR

AI and IPR ethics is an interdisciplinary field which addresses the moral, social and legal aspects of creating, using or protecting AI systems; it also seeks to establish and enforce values, principles and standards that should guide ethical behavior of those involved in AI and IPR from different perspectives while drawing on various sources as well as frameworks or methods.[82] This area is mainly concerned with balancing rights between creators' owners' users' beneficiaries' duties towards them; therefore, one needs to know what creators are supposed to do under law.[83] Another thing done within this discipline involves identification rating control risks associated with AI plus protection of people against such risk taking into account both their probable effects on individuals societies groups etcetera at large. More importantly it tries finding out how best maximize share widely distribute use benefits arising from artificial intelligence together with intellectual property rights so that human development may be advanced in all possible ways for everybody's wellbeing.[84] AI and IPR ethics is not only a theoretical and academic concern but also practical and policy-oriented. Law, philosophy, computer science, engineering, economics, sociology, psychology, education, health, media, industry, government, civil society organizations at national international levels all need to work together in addressing AI and IPR ethical issues. Furthermore; ethics of artificial intelligence and IPR must also involve the public as they are the ultimate consumers as well as beneficiaries of these technologies.[85] Ethics of AI & IPR is a living subject area because both AI systems themselves continually change along with everything around them such as laws about IP rights. Therefore, it should stay current with technological advances related to this field while being prepared for various possible outcomes arising from ethical considerations surrounding different contexts involving artificial intelligence applications vis-à-vis intellectual property protections worldwide. For example those addressed by the EU's AI Act and India's NITI Aayog AI ethics principles (updated 2024). Additionally, what might be considered morally acceptable or unacceptable in one situation may vary greatly from another, hence, there will always be need for flexibility when dealing with Ethics in relation to Artificial Intelligence plus Intellectual Property Rights based on specific circumstances.[86] E.g., India's National AI Strategy versus UNESCO's 2021 AI Ethics Recommendation; demand tailored ethical solutions to balance innovation, rights, and societal impact. The importance of the ethics related to AI and IPR cannot be overstated. This is because these aspects touch on different areas of human life and society such as autonomy, dignity, democracy, culture, education health care systems among others. It is for this reason that the subject has become very urgent in recent years with many people calling for more research into it so as to come up with solutions which can work universally without any adverse effects being witnessed in various parts of the world due to their implementation at local levels only. What needs to happen, therefore, is coming up with an ethical framework concerning both artificial intelligence technology as well intellectual property rights protection where they are seen not just tools but also means through which we express our humanity towards each other while living together peacefully.[87] E.g., UNESCO's AI Ethics Recommendation (2021) and India's NITI Aayog updates (2024).

---

[77] Mirko Farina, Xiao Yu and Andrea Lavazza, "Ethical considerations and policy interventions concerning the impact of generative AI tools in the economy and in society," 5 *AI and Ethics* 737–45 (2025).

[78] Kristie D. Prinz, "Managing the legal risks of artificial intelligence on intellectual property and confidential information." *Consulting Psychology Journal* (2025).

[79] Florian Martin-Bariteau and Teresa Scassa, *Artificial Intelligence and the Law in Canada* (LexisNexis Canada, 2023).

[80] Gobind Naidu and Vicknesh Krishnan, "Artificial Intelligence-Powered Legal Document Processing for Medical Negligence Cases: A Critical Review," 15 *International Journal of Intelligence Science* 10–55 (2025).

[81] Emily Jones, "Digital disruption: artificial intelligence and international trade policy," 39 *Oxford Review of Economic Policy* 70–84 (2023).

---

[82] Simon Chesterman, "Good models borrow, great models steal: intellectual property rights and generative AI" *Policy and Society* puae006 (2024).
[83] Mark Coeckelbergh, *AI Ethics* (The MIT Press, Cambridge (Mass.), 2020).
[84] Paula Boddington, *AI Ethics: A Textbook*, 1st ed. 2023 edition (Springer, Singapore, 2023).
[85] Bernd Carsten Stahl, Doris Schroeder and Rowena Rodrigues, *Ethics of Artificial Intelligence: Case Studies and Options for Addressing Ethical Challenges* (Springer International Publishing, Cham, 2023).

## 6. AI IN IPR MANAGEMENT

AI in IPR management is a topic that deals with the use and application of artificial intelligence systems and tools for the administration, implementation and upholding of intellectual property rights.[88] This promising field delivers significant benefits, exemplified by WIPO's AI-driven IP administration tools (expanded in 2024), enhancing efficiency and accuracy for stakeholders: creators, owners, users, regulators, and beneficiaries.[89] AI in IPR management aims to achieve the following objectives and functions:

i. Using AI systems which are capable of performing complex tasks faster than humans can do them cheaper or more accurately thus improving efficiency within such processes as searching, filing examining granting licensing enforcing intellectual property rights.

ii. Utilizing AI programs which have capacity to analyze big data sets quickly so as identify new insights trends patterns solutions with regard to patents copyrights trademarks etcetera thereby enhancing quality reliability output produced through managing these things.

iii. Making use of AI systems and tools that are able to provide user-friendly interfaces, services and platforms which can be personalized as well as making available for sharing intellectual property data and information relating to IPR management through databases e.g. patent database, literature database and case law database among others is one way of opening up access to such knowledge so as people can take advantage of it easily.

iv. Among the groups involved in IPR management there are researchers, developers, inventors and authors who need assistance in their creative work but this support should not only be limited within those fields because artificial intelligence can help spark off or boost anyone's' intellectual endeavors if used properly therefore

---

[86] Ikpenmosa Uhumuavbi, "An Adaptive Conceptualisation of Artificial Intelligence and the Law, Regulation and Ethics," 14 *Laws* 19 (2025).
[87] Thilo Hagendorff, "The Ethics of AI Ethics: An Evaluation of Guidelines," 30 *Minds and Machines* 99–120 (2020).

---

[88] Yifei Dong and Jianhua Yin, "The Application of Artificial Intelligence Technology in Intellectual Property Protection and Its Impact on the Cultural Industry," 10 *Applied Mathematics and Nonlinear Sciences* 20250005 (2025).
[89] Daniel J. Gervais, *The Human Cause* (Edward Elgar Publishing, 2022).

203

these stakeholders may also benefit from using AI systems or tools that can assist, augment or even inspire them towards new ideas.[90]

The problem with AI applied in IPRs is not only technicality but also legality and ethics thus there's need for rethinking about rights, obligations etc., concerning all parties concerned with AI-generated/AI-assisted IPRs. Who should own what? Who should be recognized as the author? What are the legal implications of AI ownership? These are some questions we ought to answer when dealing with this subject matter. Additionally it calls for identification of risks associated with its use in IPR management plus their evaluation against potential harm vis-à-vis real world effects upon different persons or communities affected by those actions taken up; Finally tries out ways that maximize benefits brought about by introduction into human societies such technologies aimed at managing IP rights taking into consideration social development needs too. Other crucial areas are adapting or updating current laws and policies in Intellectual Property Rights management to deal with AI while managing IPR taking into account new realities and challenges as well as enabling and promoting harmonization at national level among others. These concerns touch not just lawyers or scholars but also common people who live their lives under the pervasive influence of artificial intelligence on different spheres of everyday existence and social organization. Consequently, there must be a thorough investigation into what this technology means for intellectual property rights' protection; this calls for suggestions on how best do we respond legally towards these developments in ethics surrounding AI and IPRs?[91]

---

[90] Dolores Modic et al., "Innovations in intellectual property rights management: Their potential benefits and limitations," 28 *European Journal of Management and Business Economics* 189–203 (2019).
[91] Patricia Lalanda and Neus Alcolea Roig, "Ethical and Legal Challenges of Artificial Intelligence with Respect to Intellectual Property," in A. Baraybar-Fernández, S. Arrufat-Martín, *et al.* (eds.), *The AI Revolution* 63–80 (Springer Nature Switzerland, Cham, 2025).

## 7. IMPACT OF AI ON IPR POLICY

AI has a considerable bearing on IPR policy at both national and international levels. AI presents complicated challenges as well as opportunities for the development of IPR jurisprudence. This also calls for the co-ordination of standards across different regions. Some core aspects and dimensions of the influence of AI on IPR policy are:

a) The need to update existing intellectual property rights laws arises to accommodate the new realities of artificial intelligence. This includes the definition, scope, and criteria for intellectual property rights protection. It also involves the allocation and enforcement of intellectual property rights ownership and liability. Additionally, there is the need to address the balance and exceptions of intellectual property rights and obligations. For instance, UK, US, and Japan have initiated or conducted reviews, consultations, or studies on the impact of AI on IPR. They have proposed or implemented changes or amendments to their IPR laws and policies to address the questions raised by AI.[92] The United States, through the USPTO's 2024 AI patent guidance, the United Kingdom, via ongoing AI-IP consultations, and Japan, through policy reviews, are taking steps in this direction. Meanwhile, India's National AI Strategy (updated in 2024), begins integrating IPR considerations, although legislative clarity remains behind. However, a lack of consistency remains regarding the legal policy framework for AI and IPR. Further research is necessary to explore optimal and feasible solutions and recommendations for AI and IPR.

b) The need to harmonize laws across different countries is essential. This will facilitate and enhance the development and diffusion of AI. It will also help resolve conflicts that may arise from the cross-border use plus application of AI. For instance, some international organizations and forums have initiated or conducted discussions/studies on the impact of AI on IPR. They have proposed or adopted principles, guidelines, or recommendations for AI and IPR.[93] However, a lack of agreement and cooperation persists regarding international law and policy for artificial intelligence and IPR. More negotiation and collaboration are required for the progression, advancement, and implementation of common and consistent standards and practices for AI and intellectual property rights.

---

[92] Anke Moerland, "Artificial Intelligence and Intellectual Property Law" *Cambridge University Press*.

AI is a transformative force with a profound and lasting impact on IPR policy. AI facilitates the progression and transformation of intellectual property rights laws and institutions. It further necessitates the harmonization of standards and practices at both the national and transnational strata. It is paramount to embark upon a rigorous exploration of the influence of AI on IPR policy. Furthermore, it is imperative to proffer viable solutions. Also, delineate strategic recommendations to leverage the auspicious opportunities that AI avails within the domain.

## 8. CONCLUSION

In conclusion, the relentless advancement of AI across domains demands a critical re-examination of IPR frameworks. In India, the laws remain silent on AI-specific provisions. Judicial interpretations plus emerging strategies signal tentative steps toward adaptation. However, these steps lag behind global developments. This disparity accentuates an exigent necessity for juridical evolution to confront the singular challenges posed by artificial intelligence. In the realm of copyright jurisprudence, the lacuna of standardized definitions for AI-generated or AI-assisted creations becomes manifest in the United States' anthropocentric posture, the United

---

[93] Kristie D. Prinz, "Managing the legal risks of artificial intelligence on intellectual property and confidential information." *Consulting Psychology Journal* (2025).

Kingdom's provisions for computer-generated works, and the European Union's threshold for originality. This divergence engenders complexities in the adjudication of authorship, proprietorship, and the determination of originality. Patenting AI technology faces similar hurdles. India's Section 3(k) exclusions clash against AI's algorithmic core. Meanwhile, global tools like WIPO's 2024 AI-assisted patent search highlight paths forward that India has yet to embrace. Trade secrets, vital for protecting AI innovations, grapple with AI's dual role as creator and threat. This issue is exacerbated by a fragmented framework in India and international coordination gaps despite efforts of WIPO. Ethically, AI's impact is addressed by certain frameworks (like UNESCO's 2021 AI Ethics Recommendation and NITI Aayog's 2024 updates). These frameworks demand robust safeguards to balance innovation with societal good. AI's role in intellectual property rights management offers efficiency gains, as seen in WIPO's 2024 tools. Yet, it raises thorny questions of inventorship and accountability, necessitating legal and ethical recalibration. Policy-wise, the transformative force of AI calls for harmonized standards. These standards remain elusive amid WTO and OECD discussions. The integration of AI with broader goals like human rights and competition is pursued by the EU and India's nascent strategies. This confluence reveals IP rights' dual challenge: incentivizing AI-driven creativity while protecting stakeholders in an era of rapid technological flux. Ultimately, navigating this intersection requires a collaborative, interdisciplinary approach. This approach bridges legal, academic, and public spheres. India must move beyond its patchwork regime and leverage global precedents to craft adaptive laws. These laws should foster innovation without compromising ethics or equity. As AI continues to reshape intellectual creation, intellectual property rights frameworks must evolve in tandem. This evolution ensures they remain resilient and relevant in this ever-shifting landscape. This paper's exploration and recommendations aim to spark that dialogue. It urges proactive solutions to harness AI's promise while safeguarding the human ingenuity at its core.